\documentclass[twocolumn,twoside,slac]{revtex4}
\usepackage{graphicx}
\usepackage{fancyhdr}
\begin{document}

\title{Twelve Ways to Build CMS Crossings from ROOT Files}

%

\author{D. Chamont}
\author{C. Charlot}
\affiliation{  LLR, IN2P3, CNRS, Ecole Polytechnique, France}

\begin{abstract}
The simulation of CMS raw data requires the random selection of one
hundred and fifty pileup events from a very large set of files, to be
superimposed in memory to the signal event. The use of ROOT I/O for that purpose
is quite unusual: the events are not read sequentially but pseudo-randomly,
they are not processed one by one in memory but by bunches, and they do
not contain orthodox ROOT objects but many foreign objects and templates. In
this context, we have compared the performance of ROOT containers versus the
STL vectors, and the use of trees versus a direct storage of containers. The
strategy with best performances is by far the one using clones within trees,
but it stays hard to tune and very dependant on the exact use-case. The use of
STL vectors could bring more easily similar performances in a future ROOT
release.
\end{abstract}

\maketitle

\thispagestyle{fancy}


\section{Introduction\label{sec:intro}}


The CMS experiment \cite{CMS} is one of the two multipurpose experiments being under
construction to operate at the future Large Hadron Collider (LHC) at CERN.
A particularly important aspect of the CMS core software is the database system 
that will be used to handle the petabytes of data that the experiment 
will produce. The experiment has recently decided to move away from its Objectivity
based system in favor of an hybrid solution based on ROOT I/O \cite{ROOT}.
This paper describes the work done in this context so to evaluate the performances
of using the very specific ROOT classes (especially {\tt TTree} and {\tt TClonesArray})
for the data storage. In the meantime, CMS also started a more direct replacement
of the existing Objectivity implementation \cite{Bill}.


We did not choose to explore the many use-cases of CMS,
but rather focused on a single representative one
and studied the many ways to implement it with ROOT I/O.


The selected use-case is the last step of the simulation chain 
\cite{Production}: starting from the events produced by the detector simulation,
we simulate the raw data produced by the detector. Due to the high luminosity of
the LHC machine, this involves firstly the superimposition to the signal event of
a number of pileup events. The resulting crossing is then digitized, that is the
effect of the front end electronics is simulated so to produced a digitised crossing
or raw data. From the applications currently developped by CMS, this step is the
most critical from the I/O point of view: for the simulation of each raw data event
of $\sim 2MBytes$ size, one must load in memory about one hundred and fifty minimum
bias events of $\sim300 KBytes$ size. This requires huge memory and is an intensive
data reading process.
Moreover, this is an unusual use of ROOT I/O on several aspects:

\begin{itemize}

\item{} The CMS code contains C++ templates, STL containers and
it uses external packages whose classes cannot be instrumented
for ROOT. The ROOT support for templates, standard containers and external
classes is quite new and still not fully mature.

\item{} The pileup events, stored in ROOT files, are not read sequentially.
Ideally, such events should be taken from a large enough statistic,
so to produce uncorrelated sequences of pileup to be added on each signal event.
Since we have only a finite statistic of pileup events, we chose to select them
in a random fashion so to limit the possible effect of correlation between
different sequences. Nevertheless, the selection method is not fully
random (real randomness would imply a change of file for each new pileup event
and would have a strong impact on performances).

\item{} The events are not processed one by one in memory,
such as in standard analysis jobs as described in the ROOT
documentation. As explained before, we rather have at the
same time in memory all the hits from the signal and from the
many minimum bias events that we want to pile up.

\end{itemize}


In this context, we have compared the performance of four kinds
of containers (section \ref{sec:crossing-model}), combined with three
different ways (section \ref{sec:managers})
to store them in ROOT files. This leads to the twelve strategies
advertized in the title. Also worth to be mentionned, we really
focused on the huge event data to be transfered: we did not store
neither meta-data nor the links or pointers between the elements of
the events. Surely, the latter should be integrated in a later
version of the testbed. 


The section \ref{sec:testbed} gives a more detailed description of the testbed,
what we have tested and on which platform. Then, the section \ref{sec:results}
presents the most interesting results.

\section{Testbed\label{sec:testbed}}

\subsection{Main Use-Case\label{sec:use-case}}

\begin{figure*}[t]
\centering
\includegraphics{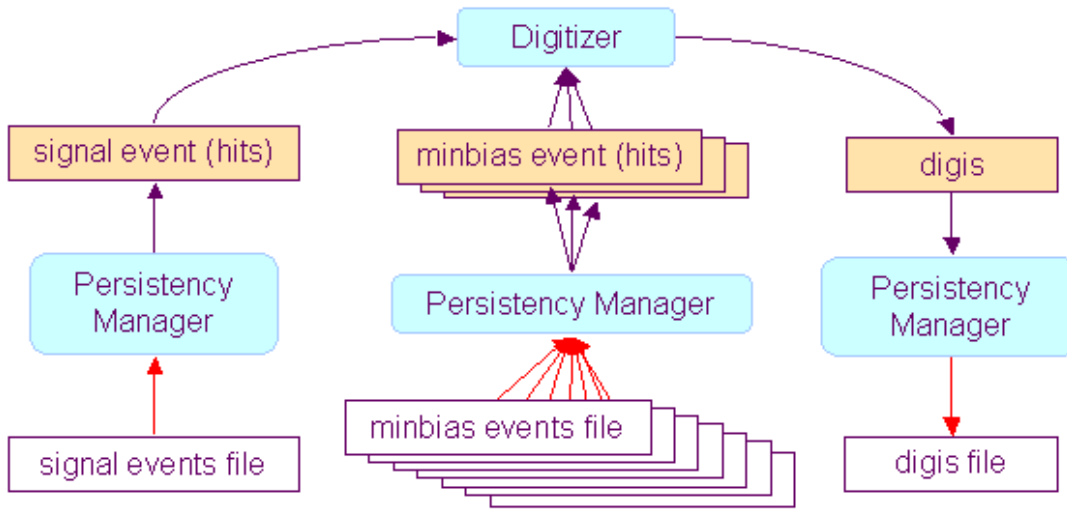}
\caption{Main Use-Case}
\label{f1}
\end{figure*}

The testbed is basically able to build crossings, as described
in figure \ref{f1}. As input, we have a file of five hundred
signal events, and one hundred files of five hundred minimum
bias events (actually, for the testbed, one hundred times the
same original file). A typical job is the production of five
hundred crossings. Building a crossing and digitizing it consists of the
following steps:

\begin{enumerate}

\item{} A persistency manager loads in memory the next signal
event from the signal file

\item{} Another persistent manager loads 153 minimum
bias events from the minimum bias files (this number corresponds to the
luminosity of $10^{34}cm^{-2}s^{-1}$). These events are
selected pseudo-randomly: we take X consecutive events, then we
do a random jump between 0 and Y, etc. X is called "burst" and
Y is called "jump". Their typical values in CMS applications are
3 and 10. Also, when loading an event, we select randomly which rank
it will have in the collection in memory.

\item{} The digitizer collects all the hits from the signal
and minimum bias events in memory, then simulates the detector
front end electronics and produces the correspondant digis.

\item{} A third persistent manager takes the digis and stores
them as the next digitized crossing in the digis output file.

\end{enumerate}

The expected bottleneck and area of interest is the loading
of the minimum bias events. Below follows the mean number
of objects and their size for such events:

\begin{itemize}
\item{} 351 instances of {\tt RtbGenParticle} (whose raw size is 46 bytes).
\item{} 584 instances of {\tt RtbSimVertex} (whose raw size is 34 bytes).
\item{} 169 instances of {\tt RtbSimTrack} (whose raw size is 38 bytes).
\item{} 3282 instances of {\tt RtbCaloHit} (whose raw size is 20 bytes).
\item{} 1871 instances of {\tt RtbTrackHit} (whose raw size ist 56 bytes).
\end{itemize}

This leads to an event size of 208 Kbytes, if we just consider the pure
data, without any adjunction for the support of persistency. As one
can see, the size is mainly dominated by the detector simulated hits. We also 
computed that the mean event size would be 392 Kbytes if each numerical
attribute would be of type {\tt double} (this information will
prove useful in section \ref{sec:matrix}).

In the results section, we will especially study the size of
the minimum bias files, and the time necessary to load 153
pseudo-random events from them.

\subsection{Crossing Data Model\label{sec:crossing-model}}

We chose to put all our persistent data in the shared hierarchy
of folders proposed by ROOT (tree of instances of {\tt TFolder}).
This permits to decouple completely the persistency mechanism
from the digitizing code.

The folder called {\tt //root/crossing/digis} is the output
"event" of the current crossing: it contains a container of instances of
{\tt RtbCaloDigis} and a container of instances of {\tt RtbTrackDigis}.

The 153 folders which we have called {\tt //root/crossing/minbias*}, plus
the single one called {\tt //root/crossing/signal} represent the
input events composing the current crossing. Each of these folders contains
a container for each kind of input event objects: {\tt RtbTrackHit},
{\tt RtbCaloHit}, {\tt RtbSimTrack}, {\tt RtbSimVertex} and {\tt RtbGenParticle}.

Each time we run the testbed, we chose a given kind of container
which is used for all the input and output data. All those
containers inherit from a common abstract class:

\begin{verbatim}
  template <class T>
  class RtbVTArray<T>
   {
    public:
    // write interface
    virtual void clear() =0 ;
    virtual void add( const T & ) =0 ;
    // read interface
    virtual UInt_t size() const =0 ;
    virtual const T & operator[]
     ( UInt_t i ) const =0 ;
   } ;
\end{verbatim}

This (simplified) class header shows that we rely on {\tt size()} and
{\tt operator[]} to read the objects of a collection. On the other side,
when creating a collection, we do not pre-allocate the size
of the container to be filled: in the CMS
code we have imported in the testbed, 
the size of the created collections is rarely known in advance. 
This write interface is not the most
efficient, but it fits the user needs and what matters most for this
testbed is the read efficiency. Four kinds of concrete containers have
been implemented, which are described below.

\subsubsection{Standard STL vector\label{sec:stl-vector}}

The class {\tt RtbStlArray<T>} wraps a standard {\tt std::vector<T>}.
The support of this kind of container is quite recent within ROOT.
The main benefit of this container is that any kind of {\tt T} can be
collected (it is not needed that {\tt T} inherits from {\tt TObject}).
The associated disadvantage is that ROOT does not apply any of its
optimizations when storing/retrieving the objects, expecially its
attribute-wise serialization.

\subsubsection{Dynamic C Array\label{sec:c-array}}

The class {\tt RtbCArray<T>} contains a simple C array dynamicaly allocated.
Each time the array is full, the size is multiplied by two and the array
reallocated.

The objects which are collected in the C array must
be instrumented with {\tt ClassDef}, otherwise ROOT I/O will not be able to save them.
Since we do not want to impose that {\tt T} is instrumented for
ROOT, we wrote a template class {\tt RtbClassDef<T>}, which inherits from {\tt T}
and is instrumented with {\tt ClassDef}. Each time we want to add an instance of
{\tt T} into the C array, we first change its type from {\tt T} to
{\tt RtbClassDef<T>}. So to evaluate the eventual cost of this process, we
also kept the possibility to compile the testbed with event data classes
directly instrumented with {\tt ClassDef} (this is done by unsetting the
macro {\tt RTB\_FOREIGN}).

The class {\tt RtbCArray<T>} should not be seen as a container to be used in a real
application. It is rather a toy container, just written
for comparison and to evaluate the performance of a good old simplistic
C array.

\subsubsection{\tt TObjArray\label{sec:objarray}}

The class {\tt RtbObjArray<T>} wraps a {\tt TObjArray}.

As for any ROOT collection, we cannot add an object to the {\tt TObjArray}
when its class does not inherit from {\tt TObject}. Again, we do
not want to impose that {\tt T} is instrumented for
ROOT, so we wrote a template class {\tt RtbObj<T>} which inherits both from
{\tt TObject} and  {\tt T}. Each time we want to add an object into
the collection, we first transform it into an instance of {\tt RtbObj<T>}.
So to evaluate the eventual cost of this process, we
also kept the possibility to compile the testbed with top event data classes
directly inheriting from {\tt TObject} (this is done by setting the
macro {\tt RTB\_TOBJECTS}).

As for {\tt RtbCArray<T>}, {\tt RtbObjArray<T>} should not be seen as a
container usable in a real application. It is here for comparison,
and should have the worst performances of all the containers, because it
is handling the objects by pointers (and not by value).

\subsubsection{TClonesArray\label{sec:clonesarray}}

The class {\tt RtbClonesArray<T>} wraps a {\tt TClonesArray}.
As for {\tt RtbObjArray<T>}, the instances of {\tt T} must
be transformed into instances of {\tt RtbObj<T>} before they are added
to the collection.

The class {\tt RtbClonesArray<T>} is the real alternative to 
{\tt RtbStlArray<T>}. Since it is based on {\tt TClonesArray}, which
is the official optimized container of ROOT, it should show the
best perfomance. The counterpart is that it is harder to use and
the collected objects must be kind of {\tt TObject}.

\subsection{Persistency Managers\label{sec:managers}}
  
The task of a persistency manager is to transfer an event from
memory (a {\tt TFolder} and its contents) to disk (an entry in
a {\tt TFile}) and vice-versa. Three flavors have been implemented,
which are described below.

\subsubsection{RtbPomKeys\label{sec:keys}}

This persistency manager implements the simplest approach,
which only uses the base ROOT I/O level and the class
{\tt TKeys}: one directly writes the {\tt TFolder} to
the {\tt TFile}, each time with a different meaningful name: the
name of the original folder plus an incrementing rank.

For example, in our main use-case, the output of the first
digitized crossing will be saved as {\tt digis0}, the
second one as {\tt digis1}, etc.


\subsubsection{RtbPomTreeMatrix\label{sec:matrix}}

The central idea of the second manager is to avoid the use of
ROOT instrumentation and dictionaries: this is achieved by
transfering the data of each container into an instance of
{\tt TMatrixD} and storing the matrix instead of the container.
More precisely, each row of the matrix is the copy of one object
from the original container, and each column corresponds to a given
attribute of the class of the objects.

In this manager, we also chose to use a {\tt TTree}. Each entry
of the tree is an event. Each branch is dedicated to one
kind of event objects, and attached to the corresponding matrix.

In this approach, we do not take profit of the ROOT persistency features,
such as the generated streamers, and we must write by ourselves the code
which transfers the data between the containers and the matrices (for
each persistent class). Also, since any number is transformed into a
{\tt Double\_t}, we expect files to be twice bigger than their normal
size.

On the other hand, we do not suffer from the ROOT parser limitations and bugs.
We are quite confident here in the data retrieved from the files, and this
manager is primarily used in the testbed so to check that the others managers
are also correctly retrieving the data.



\subsubsection{RtbPomTreeDirect\label{sec:tree}}

This third kind of manager is the ROOT recommended approach.
Each file contains a single {\tt TTree} whose entries are
the events. Similarly to the previous manager, there is one
top level branch for each kind of event object, but this
branch is directly attached to the corresponding container
in memory.

The level of split is a parameter of the testbed, but we generally
use the recommended default of 99. Concerning the size of buffers,
we rather tend to reduce them to 8000 KBytes (empirical best value).

\subsubsection{Main Use-Case}

Once a persistent manager is built, and before writing or
reading an event from a file, one must connect the manager to
a given folder in memory and to a given file.

As one can see in the main use-case (see figure \ref{f1}), there is
a manager for the signal events, and another
one for the digis, because they are always connected to the
same files and folders.

With regards to the minimum bias events, we were not able
to build a manager for each of the 153 events in memory (each
manager has some internal buffers and this would require a
large amount of memory space). Thus, we use a single manager, which must
be reconnected to a new memory folder and eventually to a new
file after reading each event. This connexion time is also
something we have closely looked at during the analysis of
the results.

\subsection{Implementation issues}

Provided one uses the option {\tt -p} of {\tt rootcint}, ROOT
has very greatly improved its support of foreign classes, templates
and std containers. It is now also possible to enforce the respect
of ANSI C++ when compiling. However, there are still some issues
with the use of ROOT that we discuss below.

\subsubsection{Documentation}

We really lack a central place where would be documented
which subset of C++ is supported in the interpreter, which subset
can be made persistent, which one can be used within a TTree, and
which one can be used with a TClonesArray within a TTree.
Since the ROOT team consider as a bug whatever is not
supported, they try to fix any such case rather than report it.
As a result, each user must rediscover by himself the unsupported
cases, when they do not do invisible damage.

\subsubsection{LinkDef}

It has proved painful to write the configuration files for the
generation of dictionaries. One must explicit all the classes which
must be parsed, and in the right order. Even with only seven top
classes to be made persistent, we felt the need to write a perl
script for the generation of the {\tt LinkDef} file. A key
cause is that when one parses a given class, one must have parsed
before all the classes of all the attributes, including each instanciated
template. 
We wonder if one could not find a way to automatize this within rootcint.

\subsubsection{Tuning of TChain branches}

Since we handle a very large number of minimum bias files with the same
internal tree, it was rather logical to use an instance of
{\tt TChain}. Actually, the fact that {\tt TChain} inherits from
{\tt TTree} is misleading. In particular, if you get the branches
and customize them, all your changes will disappear when the chain
move internally to a new file. Thus, you must detect yourself any change of
file and do again the branch customizations. In our use-case of
random events and detailed branch tuning, {\tt TChain} has finally
not proved helpful.

\subsubsection{TBranch attachment}

When one attaches a branch to a given variable, it does not
give the address of a variable, but the adress of a pointer to
a variable. This will let any C++ programmer think that he
can change later the pointer value, so to fill another
variable. This is not true! We cannot imagine any technical
reason for this, but if such an obstacle exists, the
signature of the attachment method should be changed for
the address of a variable.

%

\subsubsection{TClonesArray}

This class has really turned 
out to be hard to understand
, with a least
six size-like methods. We can understand that this comes from backward 
compatibility constraints, but still it is a problem as this class is the central
piece of the persistency service. Our proposal to have a new ROOT
collection class for the persistency (without gaps !) has not
been supported by the ROOT team.
Actually, much work is currently done for the
efficient support of {\tt std::vector<T>}, and we guess that this
class could become what we
would like to see.

\subsection{Parameters of the Testbed\label{sec:parameters}}

A few options can be set before compiling, thanks to macros
in the central header file. They have been kept as compilation
option because it was hard to make them runtime options, and
not necessarily useful. Here they are:

\begin{itemize}

\item{\tt RTB\_FOREIGN}: if set (the default), the persistent classes
are not instrumented with the macro {\tt ClassDef}, and they
will be considered as foreign classes by ROOT.

\item{\tt RTB\_TOBJECTS}: if set (not the default), the top
persistent classes inherit from {\tt TObject} and are
instrumented with {\tt ClassDef} (whatever the value of
{\tt RTB\_FOREIGN}). If not set, it implies the use of the
{\tt RtbObj<T>} when appending the objects to ROOT containers
and the use of {\tt RtbClassDef<T>} when putting them into
a dynamic C array.

\item{\tt RTB\_RESET}: if set (not the default), the
empty constructors of the persistent classes set all their
attributes to {\tt 0} (we were expecting an eventual impact
on the compression performance).

\end{itemize}

At runtime, one must choose within four kinds of containers
and three kinds of persistent managers. 
This leads to twelves base strategies. All the testbed
results will be displayed as an array of twelve cells
corresponding to these strategies (see Table \ref{t1}).

\begin{table}[t]
\begin{center}
\caption{}
\begin{tabular}{|c|c|c|c|c|}
\hline \textbf{} & \textbf{  Stl  } & \textbf{   C   } & \textbf{  Obj  } & \textbf{ Clones }
\\

\hline Keys & & & & \\

\hline Matrix & & & & \\

\hline Tree & & & & \\

\hline
\end{tabular}
\label{t1}
\end{center}
\end{table}

In the table, the third column is just there to see how bad is {\tt TObjArray},
the second column is just there to see the beavior of a good old dynamic C array,
and the second line is mainly a way to counter-check the validity of the retrieved
data. So, what matters most are the corners of the array, especially
the top left corner (Keys/Stl) which is the strategy used by CMS
for the direct replacement of its Objectivity implementation, and the
bottom right corner (Tree/Clones) which is the solution advertized
by ROOT team and evaluated in this work.


On top of the twelve strategies mentionned above, we have additional parameters,
whose value can affect the performances of these strategies differently:

\begin{itemize}

\item{Compression level}: it should slow down the writing of objects,
also slightly the reading, and reduce the size of files. A value of
{\tt 1} is expected to be the best compromise.

\item{Split level}: {\tt TTree} and ROOT containers use some sort of
attribute-wise storage mechanism. The split level is the depth of the
decomposition.

\item{Size of tree buffers}: amount of data which is read in one bunch
from the file, for a given branch.

\item{Randomness}: as described in the main use-case (see section \ref{sec:use-case}), the burst and
jump can be changed so to be close to a sequential use of minimum bias events, or
on the contrary close to a really random access pattern.

\item{Size of containers}: this parameter permits to reduce the size of events
by a given factor, so to measure the effect of this size on the performance.
The default is {\tt 1} (takes all the data). If the value is {\tt 10}, when
the input files are prepared, only 1 from 10 elements is taken.

\item{Number of crossings}: the use-case specifies that each job must build
500 crossings. This number can be lowered to see if the mean performance
is the same or to shorten the execution.

\end{itemize}

\subsection{Platform used for the tests}

All the results which are given below have been obtained
with a PC where the testbed was the only application
running. Its characteristics and software environment are
the following:

\begin{itemize}

\item{Processor}: Pentium 4, 1.8 GHz.
\item{Memory}: about 512 Mbytes.
\item{Disk}: IDE.
\item{System}: RedHat Linux 7.3 .
\item{Compiler}: gcc 3.2 .
\item{Root release}: 3.05/03 .

\end{itemize}

\section{Results\label{sec:results}}

We can hardly give here all the results we have obtained
when tuning the parameters of the testbed.
We will rather start with the best performance we have obtained,
then show and comment the effect of changing the value of some relevant
parameters.

\subsection{Best Results}

In Table \ref{t2}, you will find the size of the minimum bias files
(divided by the numbers of events, i.e. 500) and the mean time to read all
the events  of a crossing (153 minimum bias event plus a
signal event). We made the 500 crossings and used all the input data.
We also used the default compression level (1) which appeared to be always
the best compromise.

\begin{table}[t]
\begin{center}
\caption{Best Results}
\begin{tabular}{|c|c|c|c|c|}
\hline File size (Kb/event)
     & \textbf{  Stl  } & \textbf{   C   } & \textbf{Obj} & \textbf{Clones} \\
       Cpu time (s/crossing)
     & & & & \\

\hline \textbf{Keys}   & 152 & 175 & 155 & 155 \\
                       & 3.16 & 4.82 & 9.65 & 4.43 \\

\hline \textbf{Matrix} & 149 & 149 & 149 & 149 \\
                       & 2.44 & 2.85 & 3.15 & 2.72 \\

\hline \textbf{Tree}   & 153 & 176 & 156 & 54 \\
                       & 2.63 & 4.05 & 7.27 & 1.87 \\

\hline
\end{tabular}
\label{t2}
\end{center}
\end{table}

As one can expect, {\tt TObjArray} is always the worst choice
for the read performance.

Also expected, all the strategies with matrices give files of the
same size. The read time differs from one container to the other,
because of the final read step where the data is taken from the
matrices and transformed into new objects added to the containers.
It can be seen as a measurement of the efficiency of the
container {\tt add()} method.

Let's now compare {\tt std::vector} versus {\tt TClonesArray}.
Within a {\tt TTree}, {\tt TClonesArray} is by far the fastest,
and the files are incredibly smaller. The key reason here is the split 
mechanism: each attribute of each persistent class is given its own branch and buffers
(some sort of "attribute-wise" storage). This, combined with the compression of data, proves
very efficient. We do not have such performance with {\tt std::vector<>} because ROOT
does not support yet the splitting in such a case (yet one
can notice a small improvement when the instances of {\tt std::vector<>} are stored in a
{\tt TTree} rather than directly in the file). It seems that a maximum split is
always worth, even when one reads back all the branches from the tree.
With regards to the size of buffers, it appeared very complex to predict
the best value, depending on the split level, the type of object attributes and
the value of burst: it is useless to read much in advance, if there is a random
event jump coming. We proceeded empirically and finished
with a size of 8000 KBytes, largely smaller than the ROOT default.

The storage of {\tt TClonesArray} directly in {\tt TFile} (top right cell, Keys/Clones)
exhibits a rather poor performance. The reason is surely because we had to switch off
the {\tt ByPassStreamer} option, apparently buggy in such a context. As a result,
{\tt std:vector<>} is the quickest alternative when not using a tree.

\subsection{Remove compression}

\begin{table}[t]
\begin{center}
\caption{Remove compression}
\begin{tabular}{|c|c|c|c|c|}
\hline File size (Kb/event)
     & \textbf{  Stl  } & \textbf{   C   } & \textbf{Obj} & \textbf{ Clones } \\
       Cpu time (s/crossing)
     & & & & \\

\hline \textbf{Keys}   & 341 & 568 & 427 & 384 \\
                       & 1.76 & 3.00 & 8.27 & 2.95 \\

\hline \textbf{Matrix} & 400 & 400 & 400 & 400 \\
                       & 1.01 & 1.45 & 1.71 & 1.23 \\

\hline \textbf{Tree}   & 343 & 570 & 429 & 214 \\
                       & 1.53 & 2.70 & 6.17 & 1.16 \\

\hline
\end{tabular}
\label{t3}
\end{center}
\end{table}

When switching off the default compression of the data
(table \ref{t3}), one can measure how much it is efficient for
the size of file! Without compression, the read time is globally smaller,
without changing the classification of the corners.

Surprisingly enough, the strategy Matrix/Stl is really fast,
and one can wonder what it would be if its implementation was
improved. Also a surprise, but a bad one, only
the Tree/Clones and the Matrix/* strategies have file sizes
which match the predictions (see section \ref{sec:use-case}). Other
strategies have files largely too big, and we did not
fully investigate why.

\subsection{Then increase randomness and reduce data volume}

\begin{table}[t]
\begin{center}
\caption{Increase randomness}
\begin{tabular}{|c|c|c|c|c|}
\hline File size (Kb/event)
     & \textbf{  Stl  } & \textbf{   C   } & \textbf{Obj} & \textbf{ Clones } \\
       Cpu time (s/crossing)
     & & & & \\

\hline \textbf{Keys}   & 341 & 568 & 427 & 384 \\
                       & 2.20 & 3.42 & 8.63 & 3.40 \\

\hline \textbf{Matrix} & 400 & 400 & 400 & 400 \\
                       & 1.31 & 1.78 & 2.03 & 1.59 \\

\hline \textbf{Tree}   & 343 & 570 & 429 & 214 \\
                       & 2.45 & 3.71 & 6.98 & 2.71 \\

\hline
\end{tabular}
\label{t4}
\end{center}
\end{table}

Our next step has been to reduce the parameter {\tt burst} to {\tt 1} and
to increase {\tt jump} to {\tt 1000}, i.e. to take the minimum bias events
almost fully randomly.

In table \ref{t4}, one can see that increased randomness reduces more the
performance of the */Clones strategies than the */Stl, and reduces more
the performance of Tree/* than Keys/*. This makes Keys/Stl clearly the
best choice.

\begin{table}[t]
\begin{center}
\caption{Reduce data size by 10}
\begin{tabular}{|c|c|c|c|c|}
\hline File size (Kb/event)
     & \textbf{  Stl  } & \textbf{   C   } & \textbf{Obj} & \textbf{ Clones } \\
       Cpu time (s/crossing)
     & & & & \\

\hline \textbf{Keys}   & 34.9 & 54.2 & 43.6 & 39.5 \\
                       & 0.82 & 0.96 & 1.51 & 1.05 \\

\hline \textbf{Matrix} & 40.6 & 40.6 & 40.6 & 40.6 \\
                       & 0.48 & 0.51 & 0.59 & 0.52 \\

\hline \textbf{Tree}   & 35.4 & 55.0 & 44.2 & 22.7 \\
                       & 1.09 & 1.40 & 1.59 & 1.36 \\

\hline
\end{tabular}
\label{t5}
\end{center}
\end{table}

On top of that, if we reduce by a factor of ten the number of elements
we have in the events (table \ref{t5}), the effect on the specialized
ROOT classes is even worse: the strategy Tree/Clones becomes the worst
strategy (apart from Tree/C and */ObjArray) !

The lesson is quite clear: in our specific CMS use-case, the use of
{\tt TTree} and {\tt TClonesArray} is by far the most efficient strategy,
but this cannot be generalized. It highly depends on the volume of data and
the amount of randomness.

\subsection{Other Results}

Resetting the attributes to 0 in the empty constructors of the event data
does not appear to help compression. So after trying it, we went back to an implementation
where the empty contructors let undefined values in the attributes.

Unsetting {\tt RTB\_FOREIGN} or setting {\tt RTB\_TOBJECTS} has
not greatly improved the performance of the ROOT collections, so we
turned back to the use of our templated wrappers {\tt RtbClassDef<T>} and
{\tt RtbObj<T>}.

For what concerns the write cpu time of the different strategies, the
twelve strategies compares almost similarly: only the Tree/* strategies are
found slightly slower.

At last, we must confess we did not systematically build the whole set of
500 crossings that is specified by the use-case. When we did,
we always noticed that the overall performance was slightly better.

\section{Conclusion}

We have succeeded to read pseudo-random entries from a {\tt TChain} and
to dispatch them to a few hundred {\tt TFolders} (despite the fact thar the
tuning of the {\tt TChain} branches has not been straightforward).
Support for foreign classes, templates and C++ standard library has greatly
improved in the recent releases of ROOT.

The magic couple TTree/TClonesArray has proved very efficient for our
use-case, yet it requires top level {\tt TObjects} and the benefits can become
losses with smaller data volume or random access pattern. One can simply use
STL vectors and store them directly into root files. Their integration in
a {\tt TTree} is not yet as good as a {\tt TClonesArray}, but this could change
in a future release of ROOT.

If this testbed were to be improved, one major step would be
to add real pointers or {\tt TRefs} between the objects (instead of the current
indexes) and measure the impact on performance.

You can obtain the testbed source code (for linux with gmake and gcc)
by contacting the authors.

\begin{acknowledgments}

We would like to thank the ROOT team who has always quickly answered to any of
our question or bug report, and for the discussion we had concerning our results.

We are also grateful to Pascal Paganini for his contribution to
the implementation of the CMS use-case in the testbed.

\end{acknowledgments}



\end{document}